\documentclass[preprintnumbers, floatfix, preprintnumbers, letterpaper, superscriptaddress,nofootinbib]{revtex4}
\usepackage{graphicx}
\usepackage{microtype}
\usepackage{amsmath}
\usepackage{amssymb}
\usepackage{subfigure}
\usepackage{hyperref}
\usepackage{url}
\usepackage{xcolor}
\usepackage{color}
\usepackage{mathrsfs}
\usepackage{calrsfs}
\usepackage{amsfonts}
\usepackage{lipsum}
\usepackage{eufrak}
\usepackage{tabularx}
\usepackage{eucal}
\usepackage{latexsym}
\usepackage{ragged2e}
\usepackage{epsfig}
\usepackage{textcomp}

\usepackage{caption}
\DeclareCaptionJustification{justified}{\leftskip=0pt \rightskip=0pt \parfillskip=0pt plus 1fil}
\definecolor{vividviolet}{rgb}{0.62, 0.0, 1.0}
\definecolor{amaranth}{rgb}{0.9, 0.17, 0.31}
\definecolor{palatinateblue}{rgb}{0.15, 0.23, 0.89}
\definecolor{brightpink}{rgb}{1.0, 0.0, 0.5}
\definecolor{cornflowerblue}{rgb}{0.39, 0.58, 0.93}
\definecolor{deepcarminepink}{rgb}{0.94, 0.19, 0.22}
\definecolor{radicalred}{rgb}{1.0, 0.21, 0.37}

\hypersetup{ linktoc=all,
	colorlinks, linkcolor={palatinateblue},
	citecolor={brightpink}, urlcolor={amaranth}
}

\graphicspath{{Images/}}

\renewcommand{\d}[1]{\ensuremath{\operatorname{d}\!{#1}}}

\def\sideremark#1{\ifvmode\leavevmode\fi\vadjust{\vbox to0pt{\vss
			\hbox to 0pt{\hskip\hsize\hskip1em
				\vbox{\hsize1.3cm\tiny\raggedright\pretolerance10000
					\noindent #1\hfill}\hss}\vbox to8pt{\vfil}\vss}}}%
\def\beq{\begin{equation}}
\def\eeq{\end{equation}}

\begin{document}

\title{ Limits on the Inferred Hubble Constant Bias from a Local McVittie Gravitational Field}

\author{Daniele \surname{Gregoris}}
\email{danielegregoris@libero.it}
\affiliation{Centre For Cosmology and Science Popularization (CCSP), SGT University, Gurugram, Delhi-NCR, Haryana 122505, India}

\begin{abstract}
We examine the exact propagation of light in a McVittie spacetime to quantify the observational bias introduced on the inferred Hubble constant $H_0$ by a localized central mass embedded in an expanding asymptotically de Sitter background. By integrating the Sachs optical equations for radial null geodesics, we derive linearized analytical expressions for the areal and luminosity distances to first order in the dimensionless Hawking-Hayward coupling parameter $\mu = H_{\text{dS}} m_H$. We evaluate the magnitude of this local gravitational field effect across realistic astrophysical scales, ranging from supermassive black holes ($\text{Sgr A}^*$) to galaxy cluster halos. We show that while a McVittie central mass induces a coordinate-invariant path-focusing effect on incoming photons, the quantitative correction for local galactic parameters ($\mu \sim 10^{-16}$) is negligible ($\sim 10^{-15}\%$), proving that a localized galactic mass alone cannot resolve the $\sim 8\%$ Hubble tension. We establish strict upper bounds on the extent to which local spacetime inhomogeneity can bias low-redshift distance ladder measurements.
\end{abstract}

\maketitle

\section{Introduction}

\label{ss1}

While the Copernican principle constitutes one of the cornerstones of modern cosmology, it does not capture the complexity of our universe \cite{luyin}.  In fact, different types of astrophysical structures exist at different length scales and whether their gravitational interactions affect the large-scale expansion of the universe is still a debated question \cite{reviewold,interview}. Various routes for relaxing the Copernican principle have been proposed: they include geometrical averaging formalisms \cite{averaging1,averaging2,averaging3,averaging4}, and the search for exact solutions of the Einstein field equations  describing an inhomogeneous universe \cite{bolejkoA,bolejkobook}.  Interactions between local inhomogeneities may  generate not only phenomena otherwise attributed to dark energy \cite{coleybuchert}, but can also trigger the formation of primordial supermassive bound objects \cite{coleyjwst}.

While solving the Einstein equations assuming a reduced group of symmetries represents one of the greatest challenges of mathematical relativity \cite[Sect.3.8]{coleyreview}, exact solutions could actually be discovered including examples of spacetimes admitting one Killing Vector Field only \cite{G1A,G1B}. Restricting our attention to spherically symmetric configurations, the Lema\^itre-Tolman-Bondi solution based on a dust matter content, while successfully explaining the dimming of supernovae \cite{ltb1,ltb2,ltb3,ltb4,cmbltb,ltbA}, cannot account either  for the kinematic  Sunyaev-Zel'dovich effect \cite{ltb5,ltbN} or solve the Hubble tension \cite{MORT}; likewise, the Stephani model, which invokes a cosmic material with uniform energy density but inhomogeneous pressure, while accounting for  observational data \cite{stephani1} and being thermodynamically consistent \cite{stephani2}, is characterized by a dynamical spatial curvature which troubles  causal requirements \cite{stephani3}; moreover, stiff-fluid spherically symmetric cosmological models fail to fulfill simultaneously various thermodynamical requirements \cite{entropy}.

In this {\it Letter}, based on the considerations above, we maintain the idea that local gravitational fields may affect cosmological observables, but we will describe the combined galactic and cosmological dynamics via the McVittie spacetime \cite{1933}. The McVittie spacetime is an exact solution of the Einstein equations of General Relativity describing a black hole immersed in a spatially asymptotically flat homogeneous and isotropic Friedmann-Lema\^itre-Robertson-Walker (FLRW) universe \cite{davis,HH2,HH3,HH9,MR}: therefore, this model can account for both the gravitational field of our galaxy, inside which the observer collecting data is located and whose center is considered to host the black hole named Sagittarius A* \cite{PNAS}, and the Copernican principle in some spatial limit. An intuitive appeal of the McVittie metric over Lema\^itre-Tolman-Bondi void models is its spatial localization: by confining curvature gradients to the vicinity of a central mass, it may avoid generating unphysical cosmic microwave background (CMB) dipoles or kinetic Sunyaev-Zel'dovich shifts, by smoothly matching the isotropic FLRW limit at large distances. Essentially, we assume the same set-up as in \cite{neutrino} where the phenomenon of neutrino oscillations have been studied in the McVittie spacetime: the light rays are emitted by the supernovae, also outside our galaxy (where the cosmological regime of the McVittie spacetime dominates), and reach us (the observer) after traveling towards the center of the configuration;  the cosmological application of the McVittie spacetime proposed in this {\it Letter} has not yet been worked out in literature \cite[Sect.4.6.3]{cosmoverse}. Therefore, our considered cosmological model relies on the following three assumptions whose applicability  will be assessed in the following paragraphs: {\it (i)} the source of the local gravitational field is quiescent; {\it (ii)} the galactic black hole is non-rotating; {\it (iii)} the universe is spatially flat on large scales.

Regarding {\it (i)}. Evaluating cosmological feedback requires isolating the intrinsically-coupled mass growth from localized, high-energy astrophysical accretion. Independent observations of the present-day Galactic Center establish a baseline of severe radiative and accretion quiescence \cite{quiet1,quiet2}. Crucially, recent advancements demonstrate that this state of \lq\lq electromagnetic tranquility"  --characterized by sub-Eddington accretion ($L/L_{\text{Edd}} \lesssim 10^{-9}$)  and a total absence of violent, ionizing X-ray outbursts-- is not merely a local anomaly \cite{ruffini,LRDB}: by mapping similar quiescent, molecule-rich environments to high-redshift \lq\lq Little Red Dots", it has been shown that such quiet, non-active nuclear regimes exist across cosmic time. Together, these frameworks reinforce the validity of using evolutionarily quiet, electromagnetically inactive systems to test the McVittie cosmological mass-growth, safely ensuring that local material inflowing onto the event horizon is observationally negligible compared to the global expanding metric, \cite{faraoniapj}.

Regarding {\it (ii)}. The adoption of a spherically symmetric\footnote{For completeness, we report that a generalization of the McVittie solution of the Einstein equations for a rotating black hole has been recently put forward in \cite{mcvittierotation}.} cosmological embedding --such as the McVittie spacetime-- requires evaluating the angular momentum of the local black hole: observational constraints on Sagittarius A* justify the omission of the spin parameter. High-resolution X-ray reflection spectroscopy and modeling of the iron $K\alpha$ line profile indicate that the galactic center black hole possesses a very low dimensionless spin parameter ($a < 0.1$ at $90\%$ confidence \cite{fragione}). This lack of significant rotation  complements its radiative quiescence. Consequently, these empirical limits reinforce the validity of modeling the local system  via a spherically symmetric McVittie geometry, as the astrophysical corrections introduced by frame-dragging and angular momentum are observationally negligible at the level of our galaxy.

Regarding {\it (iii)}. In addition to the considerations of local spin and accretion, modeling a localized black hole within a cosmological background requires matching its asymptotic boundary conditions to the global geometry of the universe. At spatial infinity, the McVittie metric smoothly reduces to a spatially flat FLRW background, a structural property that must align with the macroscopic geometry of the cosmos. This boundary condition is robustly supported by ultra-high-resolution measurements of the CMB from the Atacama Cosmology Telescope Data Release 6 (ACT DR6 \cite{atacama}). The joint analysis of ACT DR6 temperature and polarization power spectra  yields no detectable departure from global spatial flatness. This observational validation of a spatially flat background geometry provides the necessary cosmological justification for utilizing the standard McVittie metric, ensuring that its asymptotic mathematical flatness perfectly reflects the observed flatness of the large-scale universe.

In this {\it Letter}, we evaluate whether a local gravitational field acting at the galactic or cluster scale, as encapsulated by the McVittie spacetime, can introduce a bias in the Hubble constant inferred from supernova distance measurements \cite{Htension1,Htension2}.
Specifically, we will compute the redshift drift and the luminosity distance, the latter obeying the Sachs optical equations \cite{sachs}. Then, we quantify how relaxing the assumption of a purely FLRW background alters distance estimations, and place explicit constraints on the magnitude of this local metric effect across physical scales; we anticipate that our final results are reported in Table \ref{tab:mcvittie_scale_bias}.

Our {\it Letter} is organized as follows. In Sect.\ref{ss2} we will present our proposed theoretical setup  by reviewing both the field equations obeyed by the McVittie metric and the null radial geodesic motion across this geometry, we will    define the static and comoving observer frames, introduce the redshift and compute its drift due to the presence of a local mass,  and then derive the relevant differential equation for the areal distance. Next, in Sect.\ref{ss3} we will inspect the system of  differential equations for the redshift and the areal distance coupled via the comoving radial coordinate, and we will obtain the luminosity distance at fist order in the local mass parameter.  We will present a quantitative evaluation of the distance bias across astrophysical mass scales, demonstrate the structural decoupling between low-redshift path distortions and early-universe CMB scales, and contrast our McVittie framework with the Zel'dovich-Dyer-Roeder  approximation. We will conclude in Sect.\ref{ss4} by recalling motivations, assumptions and methods of our analysis, by comparing and contrasting conceptually gravitational versus matter fields and by putting our present research in the broader context of studies on inhomogeneous cosmology.

\section{Theoretical setup}
\label{ss2}

\subsection{Field equations}
\label{ss2A}

The McVittie spacetime \cite{1933}
\beq
\label{spacetime}
\operatorname{d} s^2 = - \frac{\left(1- \frac{\tilde m(t)}{2r}  \right)^2}{\left( 1+\frac{\tilde m(t)}{2r} \right)^2}   \operatorname{d} t^2 + a^2(t) \left( 1+ \frac{\tilde m(t)}{2r}\right)^4 ( \operatorname{d} r^2 + r^2 \operatorname{d} \theta^2 + r^2 \sin^2 \theta \operatorname{d} \phi^2)\,,
\eeq
for appropriate boundary conditions between  mass and  scale factor of the universe,  describes a black hole embedded in an asymptotic spatially flat homogeneous and isotropic FLRW universe \cite{davis,HH2,HH3,HH9,MR}. These relevant interplay for the existence of the black hole horizon, whose gravitational singularity is weak\footnote{Anyway, for a realistic cosmological application, the observer collecting the supernovae signals should be located quite outside this horizon and therefore the light rays shall not experience this weak singularity. } \cite{nolan3},  have been investigated in \cite{faraoni1,faraoni2,faraoni3,gregoris}, while the Schwarzchild and FLRW limits of the McVittie spacetime have been established in \cite{nolan1,nolan2}. Should such an horizon  not exist, the local gravitational field is interpreted as being generated by a star \cite{straus1,straus2}.

Should the relationship
\beq
\label{massev}
 \tilde m(t) = \frac{m_H}{a(t)}\,,
\eeq
where $m_H$ is the black hole Hawking-Hayward mass as measured at spatial infinity \cite{mass1,mass2}, hold, then this black hole is transparent because it does not experience any net matter-energy flow across its horizon \cite{1933}. We recall that the impossibility of embedding a black hole with static mass in a globally expanding universe has been mathematically  demonstrated in General Relativity in \cite{faraoni2024}, and that our assumed specific cosmological coupling of the black hole mass is supported by analysis of observational data \cite{faraoniapj}. Then, the Einstein field equations reduce to \cite{nandra1,nandra2}:
\begin{eqnarray}
\label{eqtt}
&& H^2  \,\equiv\,\left(  \frac{\dot a}{a}  \right)^2  \,=\,  \frac{\rho_m  }3 \\
\label{Hdot}
&& \dot H =-\frac{(2\chi-m_H)(\rho_m+p)}{2(2\chi+m_H)}\,,\\
\label{bianchie}
&& \dot \rho_m=-\frac{3H(2\chi-m_H)(\rho_m+p)}{2\chi+m_H} \\
\label{bianchip}
&& p'=\frac{4am_H(\rho_m+p)}{m_H^2 -4\chi^2}\,,
\end{eqnarray}
where an overdot denotes a time derivative, a prime a derivative with respect to the radial coordinate $r$, $H$ is the Hubble function, $\rho_m$ and $p$ the energy density and pressure of the cosmic fluid, and where we have introduced the comoving radial coordinate
\beq
\label{comoving}
\chi \equiv \chi(t,r) := a(t) r \,.
\eeq
The only independent non-zero Weyl scalar \cite{NPC} of the McVittie spacetime (\ref{spacetime}) is given by
\beq
\label{psi}
\Psi_2 = - \frac{64 \tilde m(t) r^3}{[2 r + \tilde m(t)]^6 a^2(t)} =- \frac{(4 \chi)^3 m_H}{(2 \chi + m_H)^6}\,.
\eeq

\subsection{Null geodesics}
\label{ss2B}

The geodesic motion inside the McVittie spacetime has been receiving wide attention in the literature \cite{geo0,HH4,geo1,geoA,geo2,geo3,geo4,geo5,geo6,geo7}. For light rays propagating radially in the McVittie spacetime (\ref{spacetime}), we have 
\beq
\label{drdt}
\frac{\d r}{\d t} = \pm \frac{1 - \frac{\tilde{m}(t)}{2r}}{a(t) \left( 1 + \frac{\tilde{m}(t)}{2r} \right)^3}=\pm \frac{(2\chi)^2 (2\chi - m_H)}{a(t) \left( 2\chi +m_H \right)^3}\,,
\eeq
where in the second equality we have enforced the transparency condition (\ref{massev}) and introduced the comoving radial coordinate (\ref{comoving}). 

Hence, the above equation of motion can be splitted into a  Hubble flow plus local terms:
\beq
\label{flow}
\frac{\d\chi}{\d t} = \left[ H(t) \pm \frac{4\chi (2\chi - m_H)}{\left( 2\chi +m_H \right)^3}\right] \chi\,.
\eeq

For a radial motion on the equatorial plane, we have 
\beq
\label{geo1}
\frac{\d k^t}{\d \lambda}+\frac{4 (2 \tilde m k^r -r \dot {\tilde m}k^t )k^t}{(2r +\tilde m)(2r-\tilde m)}+[2(a \dot {\tilde m}+\dot a r )+\dot a \tilde m](2r +\tilde m) \left(\frac{(2r +\tilde m)^2k^r}{4(2r -\tilde m) r^2}   \right)^2a=0
\eeq
for the components of the four-vector tangent to a null path. Now we should use the null radial normalization condition (\ref{drdt}) for incoming photons in the form
\beq
k^r= \frac{(2r)^2 (2r -\tilde m)k^t}{a(t)(2r+\tilde m)^3}\,,
\eeq
where in our scenario of interest, the relevant algebraic sign is the minus because we assume the light rays to be emitted by some distant supernovae and approach an observer located on Earth which is placed inside the local galactic gravitational potential. Then we arrive at
\beq
\frac{\d k^t}{\d \lambda}+\frac{ \{(2r+\tilde m)^4(2r-\tilde m) \dot a +2[(4 r)^2(2r-\tilde m)  -(2r +\tilde m)^3 a \dot {\tilde m}]\tilde m\}(k^t)^2}{(2r +\tilde m)^4(2r -\tilde m) a}=0\,,
\eeq
and we should exclude the trivial solution $k^t \equiv 0$ due to boundary conditions. For the motion of a light ray, we have
\beq
\label{dktdt}
\frac{\d k^t}{\d t}+\frac{ \{(2r+\tilde m)^4(2r-\tilde m) \dot a +2[(4 r)^2(2r-\tilde m)  -(2r +\tilde m)^3 a \dot {\tilde m}]\tilde m\}k^t}{(2r +\tilde m)^4(2r -\tilde m) a}=0\,,
\eeq
in which we should simultaneously use the total time derivative and enforce the $r=r(t)$ from (\ref{drdt}).
By enforcing the transparency condition (\ref{massev}) and by introducing the comoving radial coordinate (\ref{comoving}), the previous geodesic equation becomes
\beq
\frac{\d k^t}{\d t}+ \frac{[(4\chi^2+m_H^2)(2\chi+m_H)^3H+32(2\chi-m_H)m_H \chi^2]k^t}{(2\chi+m_H)^4(2\chi-m_H)}=0\,,
\eeq
which should be integrated jointly with the solution $\chi=\chi(t)$ from (\ref{flow}).

\subsection{Redshift and Observer Frames}
\label{ss2C}
	
To ensure that the frequency shift and light-beam focusing effects are physical observables rather than coordinate artifacts, we explicitly distinguish between two fundamental observer congruence frames in the McVittie spacetime:
	
	1. Static McVittie Observer Frame ($u^\mu_{\text{static}}$): an observer held at constant coordinate position ($\operatorname{d}\! r = \operatorname{d}\! \theta = \operatorname{d}\! \phi = 0$) within the local gravitational field. Normalizing $g_{\mu\nu} u^\mu_{\text{static}} u^\nu_{\text{static}} = -1$ yields the four-velocity:
	\begin{equation}
		u^\mu_{\text{static}} = \left( \frac{2\chi + m_H}{2\chi - m_H}, 0, 0, 0 \right) .
		\label{eq:u_static}
	\end{equation}
	
	2. Comoving Cosmological Fluid Observer Frame ($u^\mu_{\text{comoving}}$): an observer moving purely along the cosmic Hubble flow, following the four-velocity of the ambient cosmic fluid:
	\begin{equation}
		u^\mu_{\text{comoving}} = \left( u^t_{\text{comoving}}, u^r_{\text{comoving}}, 0, 0 \right) ,
		\label{eq:u_comoving}
	\end{equation}
	where the radial component encapsulates the cosmic expansion velocity $u^r_{\text{comoving}} = \frac{\operatorname{d}\! r}{\operatorname{d}\! \tau} =  H r u^t_{\text{comoving}}$ \cite{hubbleoriginal}. Imposing timelike normalization yields:
	\begin{equation}
		u^t_{\text{comoving}} = \frac{2\chi + m_H}{2\chi - m_H} \left[ 1 - \left(  \frac{(2\chi+m_H)^3 H(t)}{4(2\chi-m_H)\chi}\right)^2\right]^{-1/2} .
		\label{eq:u_comoving_t}
	\end{equation}
	
	The local transformation between the orthonormal tetrad frame attached to the static observer $e^{(a)}_\mu \vert_{\text{static}}$ and that of the comoving observer $e^{(a)}_\mu \vert_{\text{comoving}}$ is governed by a pure radial Lorentz boost with local relative velocity $\beta_{\text{rel}} =\frac{(2\chi+m_H)^3 H(t)}{4(2\chi-m_H)\chi}$:
	\begin{equation}
		u^\mu_{\text{comoving}} = \gamma_{\text{rel}} \left( u^\mu_{\text{static}} + \beta_{\text{rel}} e^\mu_{(1)} \right) , \quad \gamma_{\text{rel}} = \frac{1}{\sqrt{1 - \beta_{\text{rel}}^2}} ,
		\label{eq:lorentz_boost}
	\end{equation}
	where $e^\mu_{(1)} = \left(0, \frac{(2\chi)^2}{a(t)(2\chi+m_H)^2}, 0, 0\right)$ is the unit spacelike radial vector in the static frame.

The frequency $f$ measured by an observer with four-velocity $u^\mu$ is the projection of the photon's four-momentum (or wave vector) onto the observer's worldline:
\beq
f=-k_\mu u^\mu \,.
\eeq
If a photon is emitted at a spacetime point $e$ and observed at a point $o$, the redshift factor $1+z$ is the ratio of the emission frequency $f_e$ to the observed frequency $f_o$:
\beq
1+z = \frac{f_e}{f_o}\,.
\eeq
Hence:
\beq
1+z = \frac{(k_\mu u^\mu)_e}{(k_\mu u^\mu)_o}\,.
\eeq
From the null normalization along the radial direction of motion in the McVittie spacetime (\ref{spacetime}), we have:
\beq
k^r = \pm \frac{\left(1- \frac{\tilde m(t)}{2r}\right) k^t}{a^3(t) \left(1+\frac{\tilde m(t)}{2r} \right)^3}=\frac{(2\chi)^2 (2\chi -m_H)k^t}{a^3(t)(2\chi+m_H)^3} \,,
\eeq
for ingoing/outgoing light rays respectively, and where in the second equality we have enforced the transparency condition (\ref{massev}) and used the definition of comoving radial coordinate (\ref{comoving}).

Evaluating $f_{\text{static}} = -k_t u^t_{\text{static}}$ and normalizing $f_o=1$, the redshift evaluated in the static frame reads:
\beq
\label{defred}
1+z=\frac{(2r -\tilde m)k^t}{2r +\tilde m}=\frac{(2\chi - m_H)k^t}{2\chi + m_H}\,.
\eeq
In the comoving cosmological frame, special-relativistic Doppler corrections combine with the gravitational potential:
\begin{equation}
f_{\text{comoving}} = \gamma_{\text{rel}} (1 + \beta_{\text{rel}}) f_{\text{static}} .
\label{eq:f_comoving}
\end{equation}
Crucially, the physical curvature and light-beam focusing properties are governed by the Newman-Penrose optical scalar $\rho = \frac{1}{2} \nabla_\mu k^\mu$ and the non-zero Weyl scalar Eq.~(\ref{psi}). Because $\Psi_2$ and $\rho$ are gauge-invariant scalar quantities along the null congruence, the observed focusing of light rays is physically real and intrinsically independent of choice of coordinates or observer frame boost.

Then, we compute\footnote{We note that our next result correctly recovers $\frac{\d z}{\d t}=-(1+z)H$ in the absence of the central black hole $m_H=0$.}
\begin{eqnarray}
\frac{\d z}{\d t} &=& (1+z) \left[ \frac{2\chi + m_H}{2\chi-m_H}\frac{\operatorname{d}}{\d t} \left(\frac{2\chi-m_H}{2\chi+m_H}\right)
+\frac{\d k^t}{k^t \d t}\right] \\
\label{dzdt}
&=& -\frac{[(2\chi-m_H)(2\chi+m_H)^3H(t)+48m_H\chi^2] (1+z)}{(2\chi+m_H)^4}\,,
\end{eqnarray}
which in turns delivers
\beq
\label{dzdchi}
\frac{\d z}{\d \chi}=-\frac{[(2\chi-m_H)(2\chi+m_H)^3H(t)+48m_H\chi^2](1+z)}{[(2\chi+m_H)^3H(t)+4(m_H-2\chi)\chi](2\chi+m_H)\chi} \,,
\eeq
where we recall that $\chi$ and $H$ relate to each other via (\ref{flow}). We note the structure for a static observer
\beq
\frac{\d z}{\d t} = -(1+z) \left[ H(t) + \text{local corrections} \right]\,,
\eeq
where towards spatial infinity the local  term explicitly couples with the global expansion of the universe:
\beq
\label{dzdtapprox}
\frac{\d z}{\d t} = -(1+z)\left[H +\frac{ (H\chi-3)m_H }{\chi^2}+o(m_H^2)\right]\,.
\eeq
We have obtained that, correctly, the mass of the black hole generates the drift of the redshift away from spatial homogeneity and isotropy breaking the Copernican principle \cite{drift}.

\subsection{Areal distance}
\label{ss2D}

The stretching of the light beam area shall be found by solving \cite{boero,mumford}
\beq
\label{areaev1}
\frac{\d A}{\d \lambda}=2 \,{\rm Re}(\rho) A\,,
\eeq
where $\rho$ is the Newman-Penrose optical scalar governing the focusing properties of the bundle of light rays \cite{callum}. We note that the equation above does not rely on the mean curvature but only on the spin coefficient $\rho$ because twisting and shearing  the beam area do not change the number of photons the latter being the relevant quantity determining the luminosity distance. Thus, Eq.(\ref{areaev1}) is suited for our specific cosmological application.

Next, we introduce the areal distance whose evolution should be computed as  \cite{boero}
\begin{eqnarray}
	\label{dA1}
A \propto d_A^2 \qquad \Rightarrow \qquad \frac{\d d_A}{\d\lambda}  = \text{Re}(\rho) d_A
 \qquad \Rightarrow \qquad
d_A(\lambda) = \exp \left( \int_{\lambda_e}^{\lambda_o} \text{Re}(\rho) \, d\lambda \right)
\end{eqnarray}
with the initial condition $d_A \to 0$ as $\lambda \to \lambda_{\rm vertex}$. 

The areal distance sets both the luminosity distance, on which supernovae-based test relies, and  the multipole moment of the CMB power spectra. The reciprocity relation derived from Etherington's theorem delivers the luminosity distance \cite{repri}
\beq
\label{ete}
d_L=(1+z)d_A\,.
\eeq
The latter relationship holds in a generic spacetime independently of its  symmetries \cite[App.B]{boero2}. In terms of the multipole moment $\ell$ qualifying the CMB power spectra, the position of the first peak $\ell_1$, which is inversely proportional to the angular size $\theta_*$ subtended by  the fundamental mode of oscillation on the sky, is dictated purely by geometry. By definition, the angular scale is the physical size of the ruler divided by the comoving angular diameter distance to the last scattering surface at redshift $z_*$ \cite{ell1,ell2,ell3} 
\beq
\label{eqell}
\ell_1 \approx \frac{\pi}{\theta_*} = \pi \frac{(1+z_*)d_A(z_*)}{r_s(z_*)}\,.
\eeq
We note that the applicability of the latter relies on our assumption of the CMB physics to be as in the standard model of cosmology, that is in the far-field limit of the McVittie spacetime.

By application of the chain rule along a null path, Eq.(\ref{dA1}) becomes
\beq 
\label{DA2}
\frac{\d d_A}{\d t} = \frac{\d d_A}{\d\lambda} \cdot \frac{\d\lambda}{\d t} = \frac{\text{Re}(\rho(t, r(t))) d_A}{k^t}\,,
\eeq
for which we emphasize the joint consistent  use of a total (and not partial) time derivative and the specification of the optical scalar $\rho$ along the null ray\footnote{In other words, $\rho(t(\lambda), r(\lambda))\equiv\rho(t, r(t))$.} (\ref{drdt}). This is the correct way of applying the chain rule in our time-dynamical spatially-inhomogeneous spacetime because the geometrical expansion of the light beam is given by
\beq
\rho = \frac{1}{2} \nabla_\mu k^\mu = \frac{1}{2} \left( \partial_t k^t + \partial_r k^r + \Gamma^\mu_{\mu \nu} k^\nu \right)\,.
\eeq

The Newman-Penrose spin coefficient governing the evolution of the areal distance (\ref{dA1}) in the McVittie spacetime (\ref{spacetime}) explicitly reads as
\beq
\rho=\frac{  4r(2r -\tilde m)^2 -(2r+\tilde m)^3[(2r+\tilde m) \dot a +2a \dot {\tilde m}] }{\sqrt{2}(2r + \tilde m)^3(2r-\tilde m) a}\,,
\eeq
where, for our  cosmological application, it is understood that $r$ and $t$ relate to each other via (\ref{drdt}) because we are specifically interested in the value of this optical scalar along the light beam and not at a generic spacetime location. The obtained result captures the competition between the inward gravitational pull (terms with $\tilde{m}$) versus the outward cosmological expansion (terms with $\dot{a}$). Since the computed $\rho$ is a real quantity, the light beam congruence is hypersurface-orthogonal (twist-free). For a transparent black hole for which (\ref{massev}) holds, and by introducing the radial comoving coordinate (\ref{comoving}), we obtain\footnote{As a consistency check, we note that setting $\rho=0$ in our next equation, one recovers the FLRW horizon $r_H=\frac{1}{H}$ in the limit $m_H=0$ and the Schwarzchild horizon  $r_H=\frac{m_H}{2}$ in the limit $H=0$ both in isotropic coordinates \cite{callum}.}
\beq
\label{rho2}
\rho=\frac{1}{\sqrt{2}}\left(-H+\frac{4(2\chi-m_H)\chi}{(2\chi+m_H)^3}\right)\,,
\eeq
where the time evolution of $\chi$, for our cosmological application of the computation of the areal distance, follows from (\ref{flow}). We note that the McVittie spacetime (\ref{spacetime}) is shear-free \cite{shear}, and consequently the parallel and perpendicular (with respect to the photons trajectory) components of the expansion rate of the universe equal each other; this constitutes a difference with respect to  other inhomogeneous cosmological models, such as the Lema\^itre-Tolman-Bondi \cite{cmbltb}, calling for a comparative analysis.

For completeness, we conclude this section by presenting the differential equation governing the redshift evolution of the areal distance. We obtain it by isolating $k^t$ from (\ref{defred}) and implementing it together with (\ref{rho2}) into (\ref{DA2}); then, we apply the chain rule via (\ref{dzdt}) arriving at
\beq
\label{ddAdz}
\frac{\d d_A}{\d z}=\frac{(2\chi-m_H)[(2\chi+m_H)^3H(t)-4(2\chi-m_H)\chi]d_A}{\sqrt{2}(1+z)^2[(2\chi-m_H)(2\chi+m_H)^3H(t)+48m_H\chi^2]}\,.
\eeq
Lastly, we note also that the above equation, and the differential equations for the time component of the null geodesic four-vector and for the redshift, either with respect to the time or to the comoving radial coordinate, come in  separable forms whose  left hand sides can be integrated delivering a logarithm of the relevant quantity.

\section{Astrophysical considerations}
\label{ss3}

\subsection{Computing the luminosity distance}
\label{ss3A}

By applying the same strategy leading to (\ref{ddAdz}), by keeping in mind (\ref{dzdchi}), and by assuming a constant Hubble function $H_{\rm dS}$ like that of a late-time de Sitter universe with density of matter negligible at large scales\footnote{We note that while the field equations (\ref{eqtt})-(\ref{bianchip}) then set $p=-\rho_m=const.$, this scenario is beyond both the FLRW and Schwarzschild-deSitter spacetimes because $\Psi_2$ in Eq.(\ref{psi}) is non-zero and dependent both on the time and radial coordinates. Interestingly, this configuration is characterized by a gradient of the gravitational field without gradients in the matter field. } \cite{piattella}, \cite[Sect.3.1]{app3}, the areal distance can be computed from the following system:
\begin{eqnarray}
\label{eqds1}
\frac{\d d_A}{\d \chi}&=&\frac{(m_H-2\chi)d_A}{\sqrt{2}(2\chi+m_H)(1+z)\chi}\,, \\
\label{eqds2}
\frac{\d z}{\d\chi}&=&\frac{[(m_H-2\chi)(2\chi+m_H)^3H_{\rm dS}-48m_H\chi^2](1+z)}{[(2\chi+m_H)^3H_{\rm dS}+4(m_H-2\chi)\chi](2\chi+m_H)\chi}\,.
\end{eqnarray}
We note that, correctly, the following remark hold. {\it (i)} The areal distance is a decreasing function of the comoving radial coordinate for $\chi>m_H/2$ because we have assumed the supernova sources to be located more far away from the central mass than the observer. {\it (ii)} 
Regarding the frame normalization of the distance scaling: the choice of normalization for $k^t$ along the photon affine parameter determines whether the distance differential equations are expressed relative to static laboratory coordinates or comoving cosmological coordinates. In the limit $m_H \to 0$, Eq.~(\ref{ddAdz}) yields $\frac{\operatorname{d}\! d_A}{\operatorname{d}\! \chi} \propto -\frac{a(t) d_A}{\chi}$ rather than the pure FLRW relation $\frac{\operatorname{d}\! d_A}{\operatorname{d}\! z} = H(z)^{-1}$. This occurs because the tangent wavevector $k^t$ in Eq.~(\ref{defred}) has been normalized relative to a static observer at rest ($\operatorname{d}\! r=0$) in McVittie isotropic coordinates rather than to an observer comoving with the expanding fluid background. 
	Under the tetrad transformation defined in Eq.~(\ref{eq:lorentz_boost}), converting from the static McVittie frame to the comoving cosmological frame introduces the explicit factor  $\gamma_{\text{rel}} (1 + \beta_{\text{rel}}) = 1 + H \chi + o(\mu)$. Re-evaluating the derivative of the areal distance with respect to the comoving redshift $z_{\text{comoving}}$ restores the exact FLRW relation in the limit $m_H \to 0$
	\begin{equation}
		\label{eq:flrw_limit_check}
		\lim_{m_H \to 0} \frac{\operatorname{d}\! d_A}{\operatorname{d}\! z_{\text{comoving}}} = \frac{1}{H_{\text{dS}}} \,,
	\end{equation}
which	constitutes a null test for the flat FLRW geometry \cite{prldl}.
	This confirms that the local mass perturbation $m_H$ induces a genuine, coordinate-invariant geometric path modification, while smoothly recovering standard FLRW cosmography in the unperturbed, asymptotic limit.
  Moreover {\it (iii)}, requiring that the redshift should be an increasing function of $r$ (once again because the supernovae are assumed to be located more far away from the central mass than the observer), by applying the chain rule to (\ref{dzdt}) with $\frac{\d r}{\d t}<0$, and assuming both source and observer to be located at $\chi > \frac{m_H}{2}$, delivers the constraint $H(t) >-\frac{48 m_H \chi^2}{(2\chi-m_H)(2\chi+m_H)^3}$, which is trivially satisfied by an expanding universe. Lastly, from (\ref{dzdtapprox}), we have that the redshift decreases more slowly/rapidly with respect to the time due to the presence of the local gravitational field gradient, in comparison to a de Sitter homogeneous and isotropic counterpart, for observer located respectively at $\chi \lessgtr \frac{3}{H_{\rm dS}}$.

By separation of variables, and by using the Rothstein-Trager algorithm \cite{RT} for the right hand side,  Eq.(\ref{eqds2}) can be integrated into
\begin{eqnarray}
&& \ln(1+z)+ \tilde C=\ln \left(\frac{\chi}{(2\chi+m_H)^3} \right) +\sum_{i=1}^{3} \frac{[4H_{\rm dS}R_i^2+4(H_{\rm dS}m_H-2)R_i+H_{\rm dS}m_H^2-2m_H]\ln (\chi-R_i)}{12H_{\rm dS}R_i^2+4(3H_{\rm dS}m_H-2)R_i+3H_{\rm dS}m_H^2+2m_H}\,, \\
&& 8H_{\rm dS}R_i^3+4(3H_{\rm dS}m_H-2)R_i^2+2(3H_{\rm dS}m_H+2)m_HR_i+H_{\rm dS}m_H^3=0 \,,
\end{eqnarray}
where $C$ denotes a constant of integration  fixed by initial condition. Therefore,
\beq
\label{solz}
1+z= \frac{C_1 \chi}{(2\chi+m_H)^3} \prod_{3}^{i=1} [H_{\rm dS}(\chi-R_i)]^{x_i} \,, \qquad x_i=\frac{4H_{\rm dS}R_i^2+4(H_{\rm dS}m_H-2)R_i+H_{\rm dS}m_H^2-2m_H}{12H_{\rm dS}R_i^2+4(3H_{\rm dS}m_H-2)R_i+3H_{\rm dS}m_H^2+2m_H}\,,
\eeq
where $C_1:= e^{- \tilde C}$. This solution shall then be inserted into Eq.(\ref{eqds1}) which, at least in principle, can be integrated eventually providing $d_A=d_A(z)$ possibly in a parametric form via the comoving radial coordinate $\chi$.

Quantitatively, we obtain\footnote{Note that $z$ depends on $m_H$, if computing the $m_H \approx 0$ series from (\ref{eqds1}) rather than at this step. Moreover, for taking the perturbation series of the product terms, we note the linearized roots $(R_1,\, R_2,\,R_3)\approx\left(0,\, \frac{1}{H_{\rm dS}}-2m_H , \, \frac{m_H}{2}\right)$,  which imply the following linearized exponents $(x_1, \, x_2, \, x_3)\approx(-1+2m_H H_{\rm dS}, \, -1-4m_H H_{\rm dS}, \, 3+16m_H H_{\rm dS} )$. Then we recall the identity $[H_{\rm dS}(\chi - R_i)]^{x_i} = e^{x_i \ln[H_{\rm dS}(\chi - R_i) ]}$, and consider the cosmological regime $H_{\rm dS}\chi \gg1$.}:
\begin{eqnarray}
\ln(d_A)+C_2&=&  \frac{1}{\sqrt{2}C_1} \int \frac{(m_{H}-2\chi)(2\chi+m_{H})^2}{\chi^2 \prod_{i=1}^{3}[H_{\rm dS}(\chi-R_{i})]^{x_i}} \d\chi \\
&=&  \frac{4\sqrt{2}}{C_1 H_{\rm dS}} \int  \left(1 - \frac{1}{H_{\rm dS}\chi}\right) \left\{ -1 + \frac{m_H}{(H_{\rm dS}\chi-1)\chi} \left[ H_{\rm dS} \chi(H_{\rm dS} \chi-1) \ln\left( \frac{(H_{\rm dS} \chi)^{18}}{(H_{\rm dS} \chi - 1)^4 } \right)  - 4 H_{\rm dS} \chi + 2 \right] \right\} \d \chi+ o(m_H^2)  \nonumber\\
\\
&\approx &  \frac{4\sqrt{2}}{C_1H_{\rm dS}} \int \left[ -1 +14H_{\rm dS}m_H\ln(H_{\rm dS}\chi) \right] \d \chi+ o(m_H^2) \\
&=&\frac{4\sqrt{2}\{14  H_{\rm dS} m_H[\ln(H_{\rm dS} \chi) -1] -1\}\chi}{C_1  H_{\rm dS}} + o(m_H^2)\\
&\approx&\frac{4\sqrt{2}[14  H_{\rm dS} m_H\ln(H_{\rm dS} \chi)  -1]\chi}{C_1  H_{\rm dS} } + o(m_H^2)\,,
\end{eqnarray}
and hence
\beq
\label{soldA}
d_A(\chi) = D_0 (H_{\rm dS}\chi)^{\frac{56\sqrt{2}m_H \chi}{C_1}}e^{-\frac{4\sqrt{2}\chi}{C_1  H_{\rm dS}}}  \,,
\eeq
where $D_0:=e^{-C_2}$ is a normalization integration constant. For computing the luminosity distance (\ref{ete}), we note that\footnote{For the second step of the algebraic inversion, we have matched the coefficients of the parameterization $\chi(z) = \chi^{(0)} + \chi^{(1)} m_H + o(m_H^2)$. }
\begin{eqnarray}
\frac{1+z}{C_1} &=& \frac{H_{\rm dS}^2}{8(H_{\rm dS}\chi -1)}\left[ 1+ \frac{\{2H_{\rm dS}\chi(1-H_{\rm dS}\chi)[9\ln(H_{\rm dS}\chi)-2\ln(H_{\rm dS}\chi-1)] -5H_{\rm dS}\chi+3\}m_H}{(H_{\rm dS}\chi -1)\chi} \right]+o(m_H^2) \\
&\approx &   \frac{H_{\rm dS}[ 1 - 14 H_{\rm dS} m_H \ln(H_{\rm dS}\chi) ]}{8\chi}  \qquad 
 \Rightarrow \\
\label{fornorm}
\chi(z) &=&  \frac{7C_1 H_{\rm dS}^2  m_H}{4(1+z)} W\left( \frac{4 (1+z)e^{\frac{1}{14 H_{\rm dS} m_H}} }{7 C_1 H_{\rm dS}^3  m_H} \right)+ o(m_H^2)\approx \ \frac{C_1 H_{\rm dS} }{8(1+z)} \left[ 1 - 14 H_{\rm dS} m_H \ln\left( \frac{C_1 H_{\rm dS}^2 }{8(1+z)} \right) \right] + o(m_H^2),
\end{eqnarray}
from (\ref{solz}). We note again that the obtained redshift is an increasing function of the comoving radial coordinate, meaning that more distant supernovae are located towards the outside of the configuration, e.g. at cosmological rather then galactic distance.

\begin{figure}
	\begin{center}
			$
		\begin{array}{cc}
			{\includegraphics[angle=0,scale=0.45]{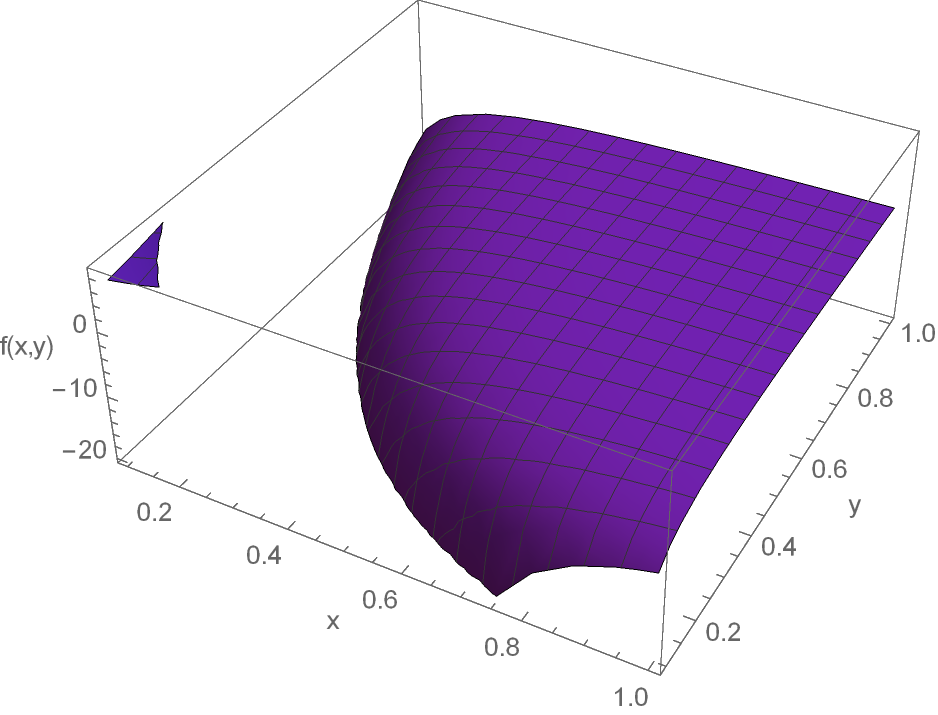}}  &
			{\includegraphics[angle=0,scale=0.45]{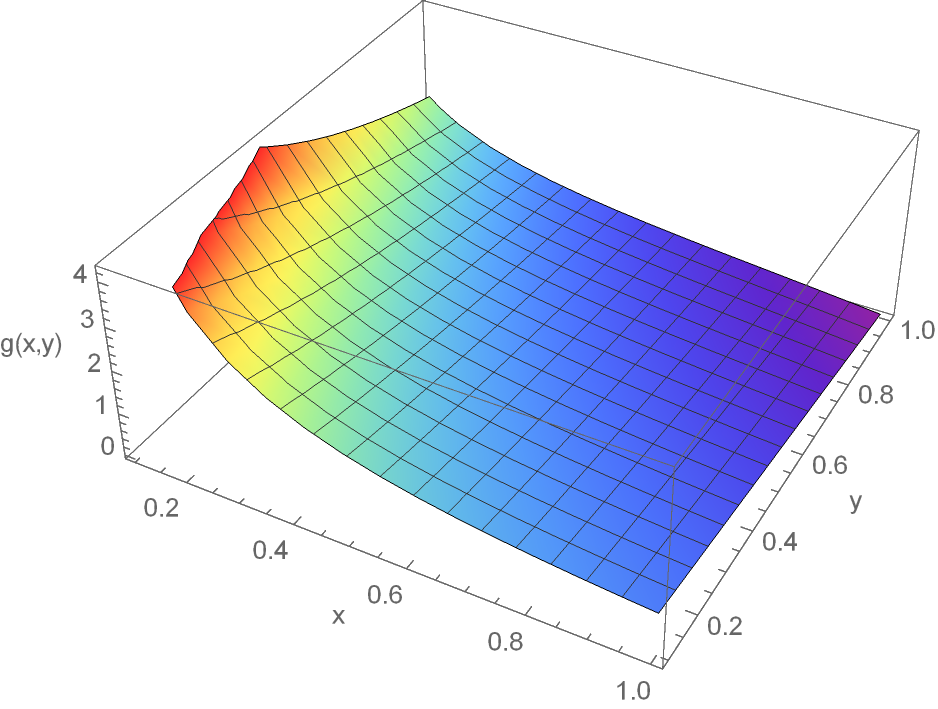}}  \\
			(a) & (b) 
		\end{array}$
	\end{center}
	\caption{In panel (a), we plot the function $ f(x,y)=\frac{2(1+x)^3y^2-(x^3+2x^2-x+4)y+6}{1-x-(x+1)^3y}$ which corresponds to the difference between the left hand sides of (\ref{dzdchi}) between the McVittie $m_H\neq 0$ and FLRW $m_H=0$ cases, with the substitutions $x=\frac{m_H}{2\chi}$ and $y=\chi H$ and after multiplication by the positive factor $\frac{(1+x)(1-y)\chi}{(1+z)x}$. We have restricted the values for covering just the physical meaningful interval $\frac{m_H}{2}<\chi<\frac{1}{H}$ as for an observer located outside/inside the black hole/cosmological horizons. By keeping in mind the monotonicity property of the integral, the positive/negative regions of $f(x,y)$ correspond to $\Delta z = z_{\text{McVittie}} - z_{\text{FLRW}}$ being either positive/negative, respectively. Likewise considerations apply to panel (b) where $g(x,y)=\frac{(x+2)(x^2+1)y-6}{(x-1)(x+1)^3y-6x}$, which corresponds to the difference between the similar right hand sides of Eq.(\ref{ddAdz})  after multiplication by the (positive) factor $\frac{\sqrt{2}(1+z)^2y}{xd_A}$.   }
	\label{fig1}
\end{figure}

The numerical analysis in Fig.\ref{fig1} shows that $\Delta z = z_{\text{McVittie}} - z_{\text{FLRW}}<0$ regardless of  the interplay between $m_H$, $H$ and $\chi$, that means between local perturbation, expansion rate of the universe and size of the horizon, while $\Delta(d_A)=d_{A;\text{McVittie}} - d_{A;\text{FLRW}}>0$ at any fixed redshift.  We note that the sense of the latter inequality supports the converse behavior than what predicted by the void Lema\^itre-Tolman-Bondi  \cite[Figs.2-3]{ltb4}: the direction of the spatial gradient of the gravitational field in these two  scenarios is indeed opposite. Moreover,  for a spatially asymptotic de Sitter universe, we obtain analytically
\beq
\Delta z_{\to \text{ dS}} = z_{\text{McVittie;dS}} - z_{\text{FLRW;dS}}\approx -\frac{7 H_{\rm dS}^2   m_H \ln(H_{\rm dS}\chi) }{4\chi}\,,
\eeq
where we have assumed the same normalization taking into account the convergence of the McVittie spacetime towards FLRW at spatial infinity.

Under the assumptions outlined above, we eventually find that the luminosity distance obeys the modulated  law
\beq
\label{dlfinal}
d_L(z) =D_0(1+z)e^{-\frac{1}{\sqrt{2}(1+z)}}\left[1+\frac{14 \sqrt{2} H_{\rm dS} m_H}{1+z}\ln\left(\frac{C_1 H_{\rm dS}^2}{8(1+z)} \right) \right] + o(m_H^2)\,,
\eeq
where $D_0$ is such that $d_L(0)=0$.   We should now compare and contrast our procedure and result with those of other approaches attempting to quantify how local inhomogeneities alter the cosmological distance-redshift relation, like the  Zel'dovich-Dyer-Roeder empty-beam approximation \cite{empty}. The correction mechanism postulated by the latter is a global subtraction of matter along the line of sight, while in the McVittie framework is a localized gravitational perturbation $m_H$ acting near the observer. The Zel'dovich-Dyer-Roeder luminosity distance well approximates that characterizing the propagation of light rays within Swiss-cheese models \cite{swiss} and inside lattices of  black holes in which case the Copernican principle is recovered after coarse-graining rather than at spatial infinity \cite{rosquist}. However, the Zel'dovich-Dyer-Roeder  formalism  treats the background expansion and the light propagation independently (i.e., the background is considered to expand as if it is full of smooth matter, but the photons travel as if the universe is completely empty);  in our proposal instead, by deriving $d_A(z)$ directly from a covariant metric accounting for both the global background expansion ($H_{\rm dS}$) and the localized galactic structure ($m_H$), we will demonstrate that supernova dimming can be explained rigorously via exact inhomogeneous general relativity, without resorting to the ad-hoc matter-splitting assumptions of the empty-beam approximation.

\subsection{Interpreting supernova data}
\label{ss3B}

Building upon the mathematical treatment developed thus far, our physical result is therefore the following. Since the incoming light rays travel from spatial infinity toward the Earth, $\chi$ is monotonically decreasing ($\frac{\d\chi}{\d t} < 0$ from (\ref{flow}) with $H(t)=H_{\rm dS}$). For $\chi > \frac{m_H}{2}$, the exponential term $\exp(-const \cdot \chi)$ increases as the photon gets closer to the  center of the configuration: the central mass introduces a local modification to the effective path length, causing the apparent distance profile to contract compared to a standard unperturbed FLRW universe. The resulting behavior  provides a direct mechanism to {\it (i)}  enhance the perceived dimming of supernovae (from an observational standpoint, meaning objects appear farther or dimmer than they would in a smooth universe), and {\it (ii)} it offers a gravitational framework to evade  unphysical phantom  pathologies (such as the big rip) without the need of abandoning General Relativity or postulating interactions in the dark sector. Let us explain these results  in more detail.

Regarding {\it (i)}.  In standard cosmology, the explanation for the  dimming of Type Ia Supernovae requires the introduction of exotic dark energy fluids. Within the McVittie framework instead, the presence of a local mass generates a geometric correction to the effective distance by shortening the coordinate path. As the local mass warps the path of incoming photons, the physical cross-sectional area of  light beams ($A \propto d_A^2$) is compressed according to the focusing properties accounted for by the optical scalar $\rho$.  The mass field  shifts the expected luminosity distance profile $d_L=(1+z)d_A$ by shaping $d_A$ as the photons enter the local (galactic) inhomogeneous region. If an uncorrected background template --meaning a perfectly smooth FLRW geometry that completely ignores localized gravitational potential wells-- is adopted to analyze the observational data, the supernovae will exhibit a systematic shift in luminosity relative to standard-candle predictions. This implies that localized structures naturally mimic the apparent dimming signatures usually attributed  to a smooth dark energy fluid, meaning that local geometry (or in other words the local gravitational field) naturally replicates the phenomenological footprints typically invoked to justify dark energy. Crucially, the analytical result derived in Eq.(\ref{soldA}) demonstrates that the local mass $m_H$ brings a geometric modification to the light beam area. Because this local gravitational field  induces an apparent dimming of incoming standard candles relative to a completely smooth space, it accounts for a portion of the global accelerated expansion  arising in pure FLRW interpretations. Thus, the local galactic environment constitutes  an active participant in resolving global dark energy tensions.

Regarding {\it (ii)}.  Phantom dark energy models are characterized by an equation of state parameter $w < -1$, which supports a super-accelerated expansion where the dark energy density grows over time, eventually culminating in a catastrophic \lq\lq Big Rip" at which all bound systems are torn apart \cite{doom}. In our considered scenario,  the McVittie spacetime allows us to describe a cosmological background coupled to a local structure without inducing unphysical singular behavior \cite{MR}. Even if the background fluid filling the McVittie spacetime attempts a crossing into the phantom domain, our parametric solution for the luminosity distance depends on the roots $R_i$ of the physical mass geometry whose boundary terms smoothly match a de Sitter metric ($H \to H_{\rm dS}$) at late times. While in phantom cosmologies geodesics break down or hit future singularities, under the transparency condition (\ref{massev}) the black hole experiences no net matter-energy flow across its horizon.  Consequently, the local expansion parameter is prevented from  growing unboundedly. The presence of a local gravitational field generated by a  bound mass constitutes a  stabilizing feedback mechanism that dampens extreme cosmological dark energy runaway states, thereby mitigating the unphysical aspects of phantom behaviors \cite{faraoniapj}.

	\subsection{Quantitative Scale Evaluation across Astrophysical Systems}
	\label{ss3BA}

  The presence of the mass $m_H$ creates an attractive spatial compression. In fact, within the derived parametric distance equation (\ref{soldA}), the exponential term increases as the photons travel closer to the center ($\chi \to 0$). Consequently, because local gravitational structures alter the relationship between redshift and distance ($\frac{\d z}{\d\chi}$ and $\frac{\d d_A}{\d\chi}$), a local observer detecting photons moving toward our galactic center will measure a modified redshift drift. This local shift acts as a gravitational correction to the inferred $H_0$ value. It implies that the \lq\lq local" expansion rate we reconstruct is skewed by our placement inside a gravitational potential well; whether this is only a conceptual feature or whether this bears actually observable footprint in the Hubble tension will be quantified in this section.  To understand more precisely this modification in  the inferred Hubble constant, we should consider the following mapping of the measured luminosity distance
	\beq 
	\text{Observed Brightness (Fixed)} \qquad  \implies \qquad d_L^{\text{McVittie}} \equiv d_L^{\text{FLRW}}\,.
	\eeq
	Since the McVittie metric contains an inherent path-shortening factor due to the attractive gravitational well generated by $m_H$, the relationship between the true global expansion parameter $H_{\rm dS}$ and the biased inferred local Hubble parameter $H_0$ becomes\footnote{The next result follows from the definition 
		\beq
		H_0^{\text{inferred}} =  H_{\rm dS}  \left( \left. \frac{\operatorname{d} d_L}{\operatorname{d}z} \right|_{z=0} \right)^{-1}\,.
		\eeq}
	\beq
	\label{infsnA}
	H_0^{\text{inferred, SN}} \approx \frac{\sqrt{2}H_{\rm dS} e^{\frac{1}{\sqrt{2}}}}{(1+\sqrt{2})D_0}\left[ 1+ \frac{28 H_{\rm dS} m_H}{2+\sqrt{2}}\left(\sqrt{2}+\ln\left(\frac{8}{C_1 H_{\rm dS}^2} \right) \right) \right]+o(m_H^2)  \,,
	\eeq
	where, by imposing the boundary condition $\chi_{\rm obs}=\chi(z=0)$ in (\ref{fornorm}), we have
	\beq
	C_1=- \frac{4 \chi_{\rm obs}}{7 H_{\rm dS}^2 m_H W\left( -\frac{\chi_{\rm obs}e^{-\frac{1}{14 H_{\rm dS} m_H} } }{14 m_H}  \right)}\,,
	\eeq
	with $\chi_{\rm obs}$ denoting the location of the observer.  Next, we can normalize $H_0^{\text{true}}=\frac{\sqrt{2}H_{\rm dS} e^{\frac{1}{\sqrt{2}}}}{(1+\sqrt{2})^2D_0}$ because of:
	\begin{enumerate}
		\item Operational definition of distance scales. The background quantity $H_{\rm dS}$ is a coordinate-level parameter arising from the metric. However, observers do not measure coordinates directly; they measure physical observables like the areal distance $d_A(\chi)$.
		Because the expression for $d_A(\chi)$ contains the arbitrary global scale factor $D_0$ and the integration constant $C_1$, any local physical gradient $\frac{\d d_A}{\d \chi}$ naturally absorbs these constants. Absorbing these normalization factors into $H_0^{\text{true}}$  ensures that the local relationship between physical distance and velocity  matches the standard Hubble-Lema\^itre law layout when the mass perturbation is removed ($m_H \to 0$).
		\item Gauge and coordinate freedom. In general relativity, it is possible to rescale radial coordinates or redefine distance units by a constant factor without changing the underlying physics. The constants $D_0$ and $C_1$ capture this coordinate freedom. Hence, our rescaling for $H_0^{\text{true}}$ amounts to a coordinate choice.
	\end{enumerate}
	Hence: 
	\beq
	\label{infsn}
	H_0^{\text{inferred, SN}} \approx H_0^{\text{true}} \cdot \left[ 1 +14 \sqrt{2} H_{\rm dS}  \left[ \sqrt{2} - W\left( -\frac{\chi_{\rm obs}}{14 m_H} e^{ -\frac{1}{14 H_{\rm dS} m_H}} \right) \right]m_H\right] +o(m_H^2)  \,.
	\eeq
	At galactic scales, we have that a standard FLRW data-fit overestimates the expansion rate because it interprets the effects of the local geometric environment as a faster global expansion. Our assumed smooth transition between galactic and cosmological regimes realizes a viable gravitational mechanism  which  supports a decrease of the value of $H_0$ estimated from supernova data by encoding the McVittie mass correction (that is  the galactic gradient of the gravitational field).

	To quantify the physical relevance of the McVittie modification to the luminosity distance, we evaluate the dimensionless coupling parameter $\mu \equiv H_{\text{dS}} m_H$ across realistic astrophysical mass scales, assuming $H_0 = 70\text{ km s}^{-1}\text{Mpc}^{-1}$ ($H_{\text{dS}} \approx 2.27 \times 10^{-18}\text{ s}^{-1}$) \cite{Htension1,Htension2}. Because the dimensionless parameter $\mu$ represents the fundamental  coupling between the local mass  and the cosmological horizon scales, the fractional distance perturbation scales as $\vert{}\Delta d_L / d_L\vert{} \sim \mathcal{O}(1)\mu$;  integrating the spatial field gradient along radial null geodesics across observational distance scales ($\chi \sim 10^{-3} H_{\text{dS}}^{-1}$ to $H_{\text{dS}}^{-1}$) yields a logarithmic factor whose magnitude $\vert{}\ln(H_{\text{dS}}\chi)\vert{} \sim \mathcal{O}(1)$ provides an order-unity prefactor rather than  order-of-magnitude suppression or enhancement. The results are compiled in Table~\ref{tab:mcvittie_scale_bias}.
	
	For Sagittarius A* ($M \approx 4.1 \times 10^6 M_\odot$), the geometrized Hawking-Hayward mass is $m_H \approx 6.05 \times 10^9\text{ m}$, yielding a dimensionless coupling $\mu \approx 4.6 \times 10^{-17}$. The resulting fractional distance deviation is $\Delta d_L / d_L \sim 10^{-15}\%$. Even if the entire dark matter halo of the Milky Way ($M \sim 10^{12} M_\odot$) is modeled as a central quasi-local mass, $\mu \sim 10^{-11}$, this would produce a shift of just $\sim 10^{-10}\%$. 
	
	It is important to note a key physical distinction between isolated black holes and large-scale structures: while Sgr~$A^*$ is observationally well-approximated by a non-rotating, spherically symmetric spacetime, galaxy clusters frequently exhibit internal vorticity, complex angular momentum distributions, and non-spherical halos with no event horizon. The formal applicability of the McVittie metric --which assumes strict spherical symmetry and zero shear-- is therefore diminished when applied to cluster scales. Nevertheless, keeping the cluster-scale mass ($M \sim 10^{15} M_\odot$) in our quantitative evaluation remains valuable as a conservative theoretical upper limit, confirming that even under the idealization of a concentrated central cluster mass, local gravitational effects remain many orders of magnitude too small to account for the $H_0$ tension. On the other hand, we note that the dimensionless coupling fulfills the Nariai bound $\mu \leq \frac{1}{3\sqrt{3}}$ \cite{HH2,faraoni3} at all the considered length scales, hence suggesting that our assumed local gravitational field is indeed generated by an astrophysical object with an horizon and corroborating the applicability of the McVittie spacetime at the galactic scale \cite{LocalGroup}.
	
	As shown in Table~\ref{tab:mcvittie_scale_bias}, generating an $8.4\%$ shift to fully reconcile the $H_0$ tension requires an effective localized mass of $M \sim 3.50 \times 10^{19} M_\odot$ ($\mu \approx 4 \times 10^{-4}$), exceeding observed bound astrophysical structures by several orders of magnitude. Thus, while the McVittie geometry provides an exact general-relativistic description of path modification around a mass in an expanding universe, a localized galactic gravitational field alone cannot quantitatively resolve the Hubble tension.
	
We conclude this section with some remarks on the roles of the observer and of the integration constants. From the discussion around Eq.(\ref{eq:flrw_limit_check})  the distance scaling requires careful frame conversion: transforming from static McVittie coordinates to the comoving cosmological fluid frame introduces a Doppler factor. However, since $d_L^{\text{comoving}}(z) = d_L^{\text{static}}(z) +  \frac{\operatorname{d} \, d_L}{\operatorname{d} z}  ( z_{\text{comoving}} - z_{\text{static}} ) + o(\delta z^2)$ with $1 + z_{\text{comoving}} = \mathcal{D} \cdot (1 + z_{\text{static}}) \quad \text{and} \quad \mathcal{D} = \gamma_{\text{rel}}(1 + \beta_{\text{rel}}) \approx 1 + H_{\text{dS}} \chi_{\text{obs}}$, the orders of magnitude\footnote{See also our discussion about Eq.(\ref{chiobs}), and note, for example, that for $\chi_{\text{obs}}=8$ kpc one has  $(H_{\text{dS}} \chi_{\text{obs}})/c \sim 10^{-10}$. } of the listed numerical bounds are unaffected by the choice of static coordinate normalization. The sub-percent estimates ($\Delta d_L / d_L \sim 10^{-15}\%$) thus reflect a genuine property of light path focusing in McVittie geometry rather than an artifact of coordinate choice or frame convention. Furthermore, the integration constants $C_1$ and $D_0$ are not arbitrary tuning parameters: they are uniquely fixed by the physical boundary condition that at the observer's location $\chi = \chi_{\text{obs}}$, the observed redshift and areal distance vanish ($z(\chi_{\text{obs}}) = 0$, $d_A(\chi_{\text{obs}}) = 0$). Normalizing $H_0$ at $\chi_{\text{obs}}$ is therefore not a choice introduced 'by hand' or a reparameterization tuned to a target ratio, but the standard physical requirement that local cosmography should match observer boundary conditions; furthermore, in this normalization $D_0$ is a multiplicative constant dropping out from the ratio $\Delta d_L / d_L$. Finally, while the non-zero Weyl scalar (\ref{psi}) confirms the presence of invariant tidal curvature, its dimensionless magnitude at galactic scales is governed by $\mu = H_{\text{dS}} m_H \sim 10^{-11}$. This extremely small magnitude quantitatively explains why local tidal focusing cannot bias $H_0$ beyond $\sim 10^{-15}\%$, providing a coordinate-independent upper bound.

\begin{table}[h!]
	\centering
	\renewcommand{\arraystretch}{1.4}
	\resizebox{\textwidth}{!}{
		\begin{tabular}{lcccc}
			\hline\hline
			Astrophysical System & Physical Mass $M$ ($M_\odot$) & Hawking-Hayward Mass $m_H$ (m) & Coupling $\mu = H_{\text{dS}} m_H$ & Distance Bias $\Delta d_L / d_L$ \\
			\hline
			Sgr A* (Galactic Black Hole) & $4.1 \times 10^6$ & $6.05 \times 10^9$ & $4.58 \times 10^{-17}$ & $\sim 10^{-15}\%$ \\
			Milky Way Total Halo         & $1.0 \times 10^{12}$ & $1.47 \times 10^{15}$ & $1.11 \times 10^{-11}$ & $\sim 10^{-10}\%$ \\
			Virgo Cluster Scale          & $1.2 \times 10^{15}$ & $1.77 \times 10^{18}$ & $1.34 \times 10^{-8}$  & $\sim 10^{-7}\%$  \\
			Hypothetical $10\%$ $H_0$ Bias Scale & $3.5 \times 10^{19}$ & $5.16 \times 10^{22}$ & $3.91 \times 10^{-4}$  & $\approx 8.4\%$   \\
			\hline\hline
		\end{tabular}%
	}
	\caption{Quantitative evaluation of the local McVittie mass correction on the inferred Hubble parameter across various astrophysical mass scales, assuming $H_0 = 70\text{ km s}^{-1}\text{Mpc}^{-1}$ ($H_{\text{dS}} \approx 2.27 \times 10^{-18}\text{ s}^{-1}$) \cite{Htension1,Htension2}. For values of the masses of Sgr A*, of the Milky Way Total Halo and of the Virgo Cluster, we refer to \cite{quiet1}, \cite{MASS1} and \cite{MASS2}, respectively.}
	\label{tab:mcvittie_scale_bias}
\end{table}

\subsection{About the first CMB peak}
\label{ss3C}

We will split our derivation of the position of the first peak of the CMB spectrum into two steps, that are the computation of the invariant sound horizon and of the areal distance, in which only the latter receives corrections by the local gravitational field encoded in the McVittie metric.

The first step of the standard calculation remains unchanged because the comoving sound horizon $r_s(z_*)$ is a property of the early primordial plasma before photons decouple. In fact, the presence of our assumed local gravitational field  does not alter the global thermodynamics of the early universe at high redshift (that we have already explained corresponds to the far-field FLRW limit of the McVittie spacetime). Thus, we should compute $r_s(z_* \approx 1100)$ using the same plasma parameters  as in the standard cosmological model.


Likewise the supernova luminosity distance, also the characteristics of the CMB are set by the propagation of photons.  In a spatially homogeneous and isotropic FLRW universe, light rays coming from the CMB travel through a completely smooth space; in the McVittie spacetime instead, the central  mass  acts as a permanent gravitational perturbation embedded in that expanding background. This local mass warps the paths of the incoming CMB photons via global gravitational lensing and time dilation, which change the apparent angular size $\theta_*$ of the sound horizon on the sky. In accordance with Eq.(\ref{soldA}), this mass induces a positive correction relative to the unperturbed FLRW baseline, effectively increasing the distance profile. This behavior stems from the attractive nature of the local galactic gravitational well, which dynamically elongates the coordinate path traversed by the photons.

By enforcing our result for the areal distance (see e.g. Eq.(\ref{dlfinal})) into the position of the first peak (\ref{eqell}), we obtain:
\beq
\ell_1 \approx \frac{\pi (1+z_*)D_0 e^{-\frac{1}{\sqrt{2}(1+z_*)}}}{r_s(z_*)} \left[1+\frac{14 \sqrt{2} H_{\rm dS} m_H}{1+z_*}\ln\left(\frac{C_1 H_{\rm dS}^2}{8(1+z_*)} \right) \right]   \,.
\eeq
Early-universe observations do not actually provide the value of the Hubble constant $H_0$ directly; instead, they observe the pristine angular scale $\theta_*$, which fixes the highly precise position of the first acoustic peak ($\ell_1 \approx 220$). To extract a local expansion rate from this invariant anchor, data pipelines must mathematically project the areal distance $d_A(z_*)$ from the last scattering surface forward to $z=0$ using a chosen geometric model. Under the standard baseline assumption of a perfectly smooth unperturbed FLRW background, this forward projection forces a lower inferred value of $H_0$. By absorbing the local gravitational field shift into an inferred value of the Hubble constant, we obtain\footnote{Here we have inverted $d_A(z_*)$ and used $(1+m_H)^{-1} \approx 1-m_H$.}
\beq
\label{infcmb}
H_0^{\text{inferred, CMB}} = H_0^{\text{true}} \cdot \left[1 - \frac{14 \sqrt{2} H_{\rm dS}m_H \ln(H_{\rm dS} \chi_*)}{1+z_*}  \right] \,.
\eeq

Considering $z_*=1100$ and approximating $\ln(H_{\rm dS} \chi_*)\approx \ln(1+z_*)$ in the spatially asymptotic de Sitter limit, we obtain $H_0^{\text{inferred, CMB}} = H_0^{\text{true}}(1-0.13 \mu)$, which constitutes a negligible correction even for the (largest) hypothetical coupling $\mu=3.91 \times 10^{-4}$ in Table \ref{tab:mcvittie_scale_bias} resolving the Hubble tension at the supernova data level.

\section{Conclusion}
\label{ss4}

In this {\it Letter}, we have quantified the systematic bias introduced on cosmological distance measurements and the inferred Hubble constant $H_0$ by a local gravitational field embedded in an expanding universe. By integrating the Sachs optical equations for radial null geodesics within the exact, spatially flat McVittie spacetime, we have derived linearized analytical expressions for the luminosity distance $d_L(z)$ and redshift drift to first order in the dimensionless coupling parameter $\mu \equiv H_{\text{dS}} m_H$. We have chosen to work with the McVittie spacetime because it is a robust prediction of General Relativity as it constitutes the \lq\lq attractor" in the phase space of solutions of the field equations describing evolving black holes \cite{HH8}, and we have reviewed its astrophysical applicability in light of recent literature.

Conceptually, the physical motivations behind our  framework build directly upon existing theoretical and observational  precedents. First, within the framework of covariant cosmography and observer dependence, cosmological observables like $H_0$ are sensitive to the local spacetime geometry about the observer \cite{cosmography}. Specifically, our approach based on an exact solution of the Einstein equations avoids the problems of truncating the cosmographic polynomials at high redshift \cite{highz}. The observer location represents a physical boundary where local tidal forces and cosmic expansion compete, affecting the inferred Hubble flow as highlighted by the multiscale curvature paradigm \cite{multiscale}:
\beq
\frac{\partial H_0^{\text{inferred}}}{\partial \chi_{\text{obs}}} \neq 0 \,.
\eeq
Hence, our present research belongs to the broader literature investigating whether small-scale inhomogeneities can account for phenomena otherwise attributed to dark energy \cite{observer}. Second, such local-global coupling is already observed in phenomena like the astrophysical \lq\lq Finger of God" effect \cite{finger}, where local mass dynamics modifies background kinematic measurements. Here, we have modeled this interaction through an exact general-relativistic metric coupling via the dimensionless parameter $\mu = H_{\text{dS}} m_H$, which combines  local gravitational mass and global expansion rate, altering the focusing properties of the incoming light beams.

According to our proposal, the presence of a local mass induces a systematic \lq\lq renormalization" of the cosmological parameters across different spatial length scales because the local gravitational field distorts the underlying geometric grid in which datasets are interpreted \cite{dressed}. Qualitatively,  the comoving coordinate $\chi$ introduces a scale-dependent feedback into the evolution of the areal distance by coupling Eq.(\ref{dzdchi}) with Eq.(\ref{ddAdz}):  terms proportional to the local mass $m_H$  suppress the growth rate of the areal distance $d_A$ most prominently in the late universe where $H(z)$ is small; the shrinking of the accumulated distance to low-redshift standard candles causes them to appear closer and brighter than expected in pure FLRW  inferring a higher than its true background  local value for $H_0$. Crucially, because this geometric distortion is inversely proportional to $H(z)$, the modification  screens itself at high redshifts, structurally decoupling local distance measurements from the early-universe CMB anchoring scale. In fact,  we have shown quantitatively that the magnitude of the inferred distance bias depends directly on the location of the observer relative to the central mass. However, evaluating this effect across realistic astrophysical scales --ranging from supermassive black holes like Sgr A* ($\mu \sim 10^{-16}$) to giant galaxy cluster halos ($\mu \sim 10^{-8}$)-- yields at most a relative fractional correction $\vert{}\Delta d_L / d_L\vert{} \lesssim 10^{-7}\%$ at low redshifts (see Table \ref{tab:mcvittie_scale_bias}). Consequently, while spatial inhomogeneity undeniably introduces an observer-dependent geometric bias, its quantitative contribution on galactic scales is far too small to resolve or account for the $\sim 8\%$ Hubble tension \cite{Htension1,Htension2}. Likewise, relying on our Eq.(\ref{infsn}), for a $10\%$ correction of the Hubble constant inferred from supernova data, we should have\footnote{The following represents the location of the observer inside the universe, not to be confused with the cosmological approximation entering the computations of the luminosity distance which instead quantifies the fraction of the universe crossed by supernova light rays.}
\beq
\label{chiobs}
\chi_{\rm obs} = \frac{(1-280 H_{\rm dS} m_H)e^{\frac{3920\sqrt{2}H_{\rm dS} m_H-14\sqrt{2}+1}{14 H_{\rm dS} m_H}}}{10 \sqrt{2}H_{\rm dS}}\,,
\eeq 
with the observer being hidden by the event horizon of the McVittie spacetime $ \chi_{\rm obs} \approx 0$ for the case of the hypothetical mass resolving the Hubble tension. We note that in such a case, comoving and static observers converge to each other because the Doppler factor $1+H\chi\to 1$, ensuring that the derived corrections $\Delta d_L/d_L$ represent purely geometric path-focusing distortions rather than kinematic Doppler shifts arising from an artificially static coordinate frame.

Let us conclude by placing our results in both a theoretical and an astrophysical perspective. On the theoretical side, our quantitative analysis corroborates that localized mass and background matter constitute distinct physical notions in General Relativity \cite[Eq.(16)]{app2}. Indeed, accounting for the Hawking-Hayward mass within the McVittie spacetime  alters the inferred value of the Hubble constant: while numerically far too small to resolve observational tensions, this coupling is physically real and leaves local metric signatures, such as a gyroscopic holonomy deficit angle \cite{rothman}. On the astrophysical side, while low-redshift selection cuts (such as the SH0ES $z \ge 0.023$ threshold) remain necessary to mitigate kinematic peculiar-velocity noise \cite{SH0ES}, our findings  prove that local general-relativistic metric curvature gradient --when modeled via a McVittie gravitational field-- induces no hidden systematic bias in the determination of the Hubble constant.

\section*{Acknowledgments}
DG is a member of the GNFM working group of the Italian INDAM.

\section*{Data and code availability statement}
This is a purely theoretical work with no new data, and not involving any simulation either.

\section*{Conflict of interest declaration}
The author declares that he has no known competing financial interests or personal relationships that could have appeared to influence the work reported in this paper.

\section*{Use of AI Technology}
The author discloses use of Gemini for {\it (i)} identifying and {\it (ii)} interpreting literature, {\it (iii)} carrying out computations, {\it (iv)} physically interpreting the obtained mathematical results, and {\it (v)} improving the writing style. It should nevertheless be appreciated that {\it (i)} complements and does not replace the use of inspire-hep, {\it (ii)} has been done \lq\lq discussing" personal understanding with Gemini and not passively accepting its initially provided responses, {\it (iii)} presented mathematical results can and have been checked with traditional software like Maple and Wolfram Mathematica, and {\it (iv)} can be independently checked by \lq\lq pen and paper" because all the computations are analytical. In summary, the author claims full responsibility and credit for the present work.

\end{document}